\documentclass[twoside]{article}
\usepackage{fleqn,espcrc2}
\usepackage[dvips]{graphicx}
\usepackage[figuresright]{rotating}
\usepackage{epsf}

\usepackage[dvips]{graphicx}
\usepackage[figuresright]{rotating}
\usepackage{epsf}
\usepackage{amssymb}
\usepackage{amssymb}

\newcommand{\cO}{{\cal O}}
\newcommand{\ket}{\,\rangle}
\newcommand{\bra}{\langle \,}

\title{Quantum loops in the Resonance Chiral Theory: \\
improving the vector form factor
\thanks{Talk given at the International Workshop $e^+e^-$ Collisions from $\phi$ to $J/\psi$, 27th February--2nd March (2006), Novosibirsk (Russia). IFIC/06-19; FTUV/06-30-05 report.}
}
\author{I.~Rosell\address{Departament de F\'{\i}sica Te\`orica, IFIC,
CSIC-Universitat de Val\`encia, \\ Apt. Correus 22085, E-46071 Val\`encia,
Spain}} 
       
\begin{document}

\begin{abstract}
Quantum loops in the Resonance Chiral Theory are needed to improve the implementation of non-perturbative QCD.  
Furthermore, the one-loop computations can predict chiral low-energy couplings at next-to-leading order, a very appealing task. We present a first calculation of the vector form factor of the pion at subleading order in the $1/N_C$ expansion. From the analysis of the result at large energies we justify the requirement of considering short-distance constraints from form factors with resonances in the final state. The long-distance limit of our results allows to get a next-to-leading order estimate of $\ell_6$.
\vspace{1pc}
\end{abstract}

\maketitle

\section{Introduction}

The non-perturbative nature of Quantum Chromodynamics (QCD) at long distances prevents from using directly the QCD lagrangian at low energies. An effective field theory formalism allows an approach to this issue~\cite{EFT}.

At energies between the $\rho$ mass and $2$ GeV there are some problems that make more difficult this procedure. The absence of a natural expansion parameter and the presence of many resonances without a mass gap hinder a formal effective development. Notwithstanding, the $1/N_C$ expansion turns out to be a suitable tool to carry out this aim~\cite{N_C}. 

In the limit of a large number of colours QCD can be described by the tree diagrams of an effective lagrangian with local operators and an infinite number of meson fields, being loop corrections suppressed in the $1/N_C$ expansion. By taking into account this point of view the success of the Vector Meson Dominance (VMD) in the sixties and seventies is explained. 

Resonance Chiral Theory (R$\chi$T) is an effective theory of QCD at intermediate energies that can be understood following these ideas~\cite{RChTa,RChTb,RChTc}. The terms that can be constructed observing the symmetries of the underlying theory, QCD, and the very low-energy effective theory, Chiral Perturbation Theory ($\chi$PT), are considered.

The model dependence of R$\chi$T comes from the cut in the tower of resonances. In any case this is supported by the phenomenology and one assumes that contributions from higher resonances are suppressed by their masses. In Ref.~\cite{RChTa} only contributions from the lightest meson with any given quantum numbers were taken into account, the so-called Single Resonance Approximation.

As in any effective theory description, the matching with the high-energy theory is crucial, i.e. R$\chi$T should recover the short-distance behaviour of QCD. This excludes interactions with large number of derivatives and gives a lot of constraints between the couplings.

Quantum loops in the Resonance Chiral Theory are needed to improve predictions of non-perturbative QCD. In fact, there are many observables where the predictions for the hadronic contributions should get better to be able to distinguish New Physics from Standard Model.

The prediction of the couplings of $\chi$PT is one of the most important profits of R$\chi$T. Nevertheless, with only tree-level calculations only leading-order predictions can be found. 

It is important to stress that the saturation of the couplings of the $\cO(p^4)$ $\chi$PT lagrangian by the resonance exchange~\cite{RChTa}, which allows their prediction within our effective approach, is fully understood in a large-$N_C$ scenario.

As a first step in the next-to-leading order calculations within R$\chi$T, we have studied the pion vector form factor (VFF)~\cite{nosaltres}. This physical amplitude has been chosen because of its simplicity and phenomenological importance, keep in mind for instance the $e^+e^-$ colliders. This observable is defined through the two pseudo-Goldstone bosons matrix element of the vector current:
\begin{equation}
\bra \pi^+ \pi^- | (\overline{u} \gamma^\mu u -\overline{d} \gamma^\mu d)/2  | 0 \ket = \mathcal{F}(q^2) (p_+-p_-)^\mu  ,
\end{equation}
where $p_+$ and $p_-$ are the momentum of the $\pi^+$ and $\pi^-$ respectively and $q^\mu=(p_++p_-)^\mu$.


The calculation at hand has been made simple by working in the two flavours theory and taking the chiral limit. Therefore, considering the results at $N_C \rightarrow \infty$, a chiral $U(2)_L \otimes U(2)_R$ will be assumed.
 
\section{The lagrangian of Resonance Chiral Theory}

As a first approach to the one-loop calculations of R$\chi$T only interaction terms up to one resonance field have been considered, that is, we use the lagrangian of Ref.~\cite{RChTa}, where the Single Resonance Approximation is followed. It contains a first piece without resonances, the $\cO(p^2)$ $\chi$PT lagrangian,
\begin{eqnarray}
\mathcal{L}_{2\chi} &=& \frac{F^2}{4} \bra u_\mu u^\mu  +  \chi_+ \ket \, ,
\end{eqnarray}
the kinetic resonance lagrangian,
\begin{eqnarray}
\mathcal{L}_{2Z}& =&- {1\over 2} \bra \nabla^\lambda R_{\lambda\mu} \nabla_\nu R^{\nu\mu} -{1\over 2} M^2_R \, R_{\mu\nu} R^{\mu\nu}\ket \, , \nonumber \\
\mathcal{L}_{2Z} & =&{1\over 2} \bra \nabla^\mu R\,\nabla_\mu R - M^2_R\, R^2\ket \, ,
\end{eqnarray}
where the first expression reads for the $R=V,A$ case and the second one for the $R=S,P$, and $\cO(p^2)$ interactions linear in the resonance fields:
\begin{eqnarray}
\mathcal{L}_{2V}[V] & = &  \frac{F_V}{ 2\sqrt{2}}  \bra V_{\mu\nu} f_+^{\mu\nu}\ket + \frac{i\, G_V}{\sqrt{2}} \bra V_{\mu\nu}  u^\mu   u^\nu  \ket \, ,\nonumber\\
\mathcal{L}_{2A}[A] & = & \frac{F_A}{ 2\sqrt{2}}  \bra A_{\mu\nu} f_-^{\mu\nu} \ket\,  ,\nonumber \\
\mathcal{L}_{2S}[S]  & = &  c_d  \bra S \,u_\mu u^\mu \ket  + c_m  \bra S \,\chi_+ \ket \,  ,\phantom{\frac{1}{2}}\nonumber \\
\mathcal{L}_{2P}[P] & = &  i\, d_m  \bra P \,\chi_-\ket \,  .\phantom{\frac{1}{2}}
\end{eqnarray}
Notice that the notation of Ref.~\cite{RChTa,nosaltres} is kept and in the chiral limit and neglecting external scalar or pseudoscalars sources $\chi_\pm=0$.

Our resonance chiral lagrangian follows the $N_C$ counting rules: the different fields, masses and momenta are of $\cO(1)$ in the $1/N_C$ expansion, whereas taking into account the interaction terms one can check that $F,F_V,G_V,F_A,c_d,c_m$ and $d_m$ are of $\cO(\sqrt{N_C})$. At $N_C\rightarrow \infty$ only operators with one trace are allowed and the $U(2)_R \otimes U(2)_L$ symmetry is followed in the two flavours effective case.

In the Introduction the necessity of fulfilling the QCD behaviour at large energies has been stressed. Using different matrix elements, all the hadronic parameters introduced previously are fixed in terms of the pion decay constant $F$ and the two masses of the vector and scalar multiplet, $M_V$ and $M_S$~\cite{polychromatic}, $F_V/\sqrt{2} = \sqrt{2}\, G_V = F_A = 2 c_m = 2 c_d = 2 \sqrt{2}\, d_m = F$ for the couplings and $M_A = \sqrt{2}\, M_V\,, \,M_P \simeq \sqrt{2}\, M_S$ in the case of the resonance masses.

Through the renormalization procedure different counterterms are needed in order to absorb the ultraviolet divergences. The minimal set of chiral structures have been included and we expect their couplings to be subleading in the $1/N_C$ expansion, since they are associated with quantum loops corrections. Assuming this fact, at next-to-leading order these counterterms contribute only at tree-level. Therefore the lowest-order equations of motion can be used to reduce the number of operators. In Ref.~\cite{nosaltres} it is explained how only the leading lagrangian and subleading pieces without resonance fields are needed in the renormalization of the vector form factor of the pion, once the equations of motion are used,
\begin{eqnarray}
\widetilde{\mathcal{L}}_{4\chi} &=& \frac{i\,\widetilde{\ell}_6}{4} \bra f_+^{\mu\nu} \left[ u_\mu , u_\nu \right]  \ket  - \widetilde{\ell}_{12}\bra \nabla^\mu u_\mu \nabla^\nu u_\nu\ket \,  , \nonumber \\
\widetilde{\mathcal{L}}_{6\chi}& = & i \,\widetilde{c}_{51} \bra \nabla^\rho f_+^{\mu\nu} [h_{\mu\rho},u_\nu] \ket  + \nonumber \phantom{\frac{1}{2}}\\&&\quad \,+ i \,\widetilde{c}_{53} \bra  \nabla_\mu f_+^{\mu\nu} [h_{\nu\rho},u^\rho] \ket \, ,\phantom{\frac{1}{2}}\label{subleading}
\end{eqnarray} 
where the tilde is used to denote the R$\chi$T couplings, unlike the $\chi$PT ones.

Including the subleading lagrangian of Eq.~(\ref{subleading}), the tree-level calculation of the pion vector form factor reads
\begin{equation}
\mathcal{F} (q^2) \,=\, 1  + {F_V\, G_V\over F^2} {q^2\over M_V^2-q^2} - \widetilde{\ell}_6 {q^2\over F^2} + \widetilde{r}^{\phantom{\, r}}_{V2}\, {q^4\over F^4}\, ,
\end{equation}
where $\widetilde{r}^{\phantom{\, r}}_{V2}\equiv 4 F^2\,(\widetilde{c}_{53}- \widetilde{c}_{51})$ has been defined.
Taking into account the Brodsky-Lepage behaviour of the two-body form factors, i.e. they should vanish at large momentum transfer, the confirmation of the next-to-leading nature of the new couplings is found, that is $\widetilde{\ell}_6=0$ and $\widetilde{r}^{\phantom{\, r}}_{V2} = 0$ at leading order in the $1/N_C$ expansion, as we have argued before. Thus, one can establish a well defined counting in powers of $1/N_C$ to organize the calculation.



\section{Resonance saturation at next-to-leading order}
The low-energy limit of our result allows us to investigate the resonance saturation at next-to-leading order in $1/N_C$, because of the rigorous control of the renormalization scale dependences. It is found that
\begin{eqnarray}
\ell_6^{\,r}  &=&  - \frac{F_V G_V^{\,r}}{M_V^2}  + \widetilde{\ell}_6^{\,r} + \frac{1}{16\pi^2}\left[ \frac{4}{3}\ln{\frac{M_V^2}{\mu^2}} \right.\nonumber \\
&&  - \frac{1}{2}\ln{\frac{M_A^2}{\mu^2}} + \frac{1}{6}\ln{\frac{M_P^2}{\mu^2}} - \frac{M_S^2}{M_V^2}\ln{\frac{M_S^2}{\mu^2}} \nonumber \\ 
&&\left.+ \frac{11}{18} + \frac{M_S^2}{2 M_V^2} \right]\, , \label{l6}\\
r^{\, r}_{V2} &=&\frac{F^2 F_V G_V^{\,r}}{M_V^4} + \widetilde{r}^{\, r}_{V2} + \frac{2 F^4}{M_V^2} \left[\hat{X}-X_Z^r(\mu)\right] \nonumber\\
&&+\frac{F^2}{96\pi^2}\left\{\left(6\frac{M_S^2}{M_V^4}+\frac{1}{2 M_V^2} -\frac{1}{2 M_S^2}\right)\right. \times \nonumber \\
&&\left. \times \ln{\frac{M_S^2}{\mu^2}} -\frac{9}{M_V^2}\ln{\frac{M_V^2}{\mu^2}}-\frac{1}{M_A^2}\ln{\frac{M_A^2}{\mu^2}} \right.\nonumber \\&&\left. -\frac{167}{60 M_V^2}- \frac{17}{10 M_A^2} - \frac{3 M_S^2}{M_V^4} + \frac{17}{20 M_S^2} \right.\nonumber \\
&& \left. +\frac{1}{10 M_P^2}\right\}\, ,\label{rv}
\end{eqnarray}
where new couplings and the different runnings are shown in Ref.~\cite{nosaltres}. From these expressions the known lowest-order predictions for the two $\chi$PT couplings can be checked, $\ell_6=-F^2/M_V^2$ and $r^{\, \phantom{r}}_{V2}=F^4/M_V^4$. It is important to remark the different running between the $\chi$PT and the R$\chi$T case.

The $\chi$PT couplings $\ell_6$ and $r^{\phantom{\, r}}_{V2}$ have been phenomenologically extracted from a fit to the vector form factor data at low momenta~\cite{VFF_ChPT}, 
\begin{eqnarray}
\bar \ell_6 \equiv \frac{32\pi^2}{\gamma_{\strut\ell_6}}\ell_6^{\,r}(\mu) - \ln{\frac{m_\pi^2}{\mu^2}} = 16.0\pm 0.5\pm 0.7\,,\nonumber \\
r^{\, r}_{V2}(M_\rho)=(1.6\pm 0.5)\cdot 10^{-4}.\qquad\qquad\qquad
\end{eqnarray}
Considering these values in Eqs.~(\ref{l6}) and (\ref{rv}), one can estimate the corresponding scale-invariant combinations of subleading couplings in R$\chi$T, $\hat{l}_6$ and $\hat{r}^{\phantom{\, r}}_{V2}$~\cite{nosaltres}. Taking $F=92.4$~MeV, $M_V = 770$~MeV and $M_S = 1$~GeV, one gets $\hat{\ell}_6 = (-0.2\pm 0.9)\cdot 10^{-3}$ and $\hat{r}^{\phantom{\, r}}_{V2}= (-0.2\pm 0.5)\cdot 10^{-4}$, while a larger value of the scalar resonance mass $M_S = 1.4$~GeV shifts the $\cO(p^4)$ coupling to $\hat{l}_6 = (-0.9\pm 0.9)\cdot 10^{-3}$, without affecting $\hat{r}^{\phantom{\, r}}_{V2}$ at the quoted level of accuracy. Compare these values with the large--$N_C$ predictions $\ell_6|_{LO} = -F^2/M_V^2 = -0.014$ and $r^{\phantom{\, r}}_{V2}|_{LO} =F^4/M_V^4 = 2.1\cdot 10^{-4}$. In other words, the resonance saturation is proved to work accurately at next-to-leading order in the $1/N_C$ expansion.

\section{High-energy behaviour}

The ultraviolet behaviour of our calculation can be analyzed: it has a wrong behaviour $\mathcal{F}(q^2)\sim q^4 \ln (-q^2/\mu^2)$, which cannot be eliminated with a local contribution. The problem originates in the two-resonance cut. Actually, it is not surprising taking into account that only the lagrangian of Ref.~\cite{RChTa} has been considered, since only interaction terms with a resonance field are incorporated. Notice that from large-$N_C$ arguments there is no limit in the number of resonances needed to build the terms, while we only incorporate some bilinear interactions trough the kinetic pieces. It is clear that there are many more terms with two resonances that could be added to our lagrangian~\cite{RChTc}. 

It seems appealing to conjecture that these new pieces with more than one resonance field, in this case terms with two or three resonances, give contributions to our result, should combine with the present contributions to generate the expected suppression at $q^2 \rightarrow \infty$.

From a more systematic and general point of view, this problem has been recently examined~\cite{nosaltres2}. All the two-body form factors that can be found in the even-intrinsic-parity sector of Resonance Chiral Theory in the Single Resonance Approximation have been analyzed in the spirit of the correlators at next-to-leading order in the $1/N_C$ expansion. In Ref.~\cite{nosaltres2} we have justified the necessity of considering well-behaved form factors with resonances in the final state before performing one-loop calculations. Furthermore, it has been observed that no local $\chi$PT counterterms are required in the renormalization of some form factors and correlators. 
This is very useful, since calculations of correlators or form factors at one-loop level provide then a clear next-to-leading order prediction of the related $\chi$PT couplings. 
In Ref.~\cite{nosaltres2} we give a subleading prediction of $L_8$, where the scale dependence is under control.

\section{Summary}

We have performed a calculation of the vector form factor of the pion at next-to-leading order in the $1/N_C$ expansion, within the framework of Resonance Chiral Theory. 
\begin{enumerate}
\item The low-energy limit of our result allows to get a next-to-leading order prediction of different $\chi$PT couplings. Thus, we have been able to check the successful resonance saturation at next-to-leading order.
\item From the bad behaviour of the pion vector form factor at large energies many conclusions have followed. The requirement of new terms that contribute to the process, independently from the number of resonance fields that contain, is evident. In fact, we have justified the necessity of considering high-energy well-behaved form factors with resonances in the final state before performing one-loop calculations~\cite{nosaltres2}. This is very important, since until now demanding that two-body form factors of hadronic currents vanish at high energies was controversial in the case of resonances in the final state.
\end{enumerate}

This work is supposed to be a first effort in the calculations at one-loop level within Resonance Chiral Theory. More investigations have been done following this aim~\cite{oneloop}.

\vspace{0.4cm}

\noindent {\bf Acknowledgments}  \\
We wish to thank the organizers of the International Workshop $e^+e^-$ Collisions from $\phi$ to $J/\psi$ for the useful and pleasant congress. I.R. is supported by a FPU scholarship of the Spanish MEC. This work has been supported in part by the EU HPRN-CT2002-00311 (EURIDICE), by MEC (Spain) under grant FPA2004-00996 and by Generalitat Valenciana under grants GRUPOS03/013, GV04B-594 and GV05/015.


\vspace*{-0.1cm} 
\begin{thebibliography}{99}
\vspace*{-0.2cm}

\bibitem{EFT}
  A.~V.~Manohar,
  arXiv:hep-ph/9606222;\\
  A.~Pich,
  arXiv:hep-ph/9806303.

\bibitem{N_C}
  G.~'t Hooft,
  Nucl.\ Phys.\ B {\bf 72} (1974) 461;\\
  E.~Witten,
  Nucl.\ Phys.\ B {\bf 160} (1979) 57.

\bibitem{RChTa}
  G.~Ecker, J.~Gasser, A.~Pich and E.~de Rafael,
  Nucl.\ Phys.\ B {\bf 321} (1989) 311.

\bibitem{RChTb}
  G.~Ecker, J.~Gasser, H.~Leutwyler, A.~Pich and E.~de Rafael,
  Phys.\ Lett.\ B {\bf 223} (1989) 425.

\bibitem{RChTc}
  V.~Cirigliano, G.~Ecker, M.~Eidemuller, R.~Kaiser, A.~Pich and J.~Portoles,
  arXiv:hep-ph/0603205.

\bibitem{nosaltres}
  I.~Rosell, J.~J.~Sanz-Cillero and A.~Pich,
  JHEP {\bf 0408} (2004) 042.

\bibitem{polychromatic}
  A.~Pich,
  arXiv:hep-ph/0205030.

\bibitem{VFF_ChPT}
  J.~Bijnens, G.~Colangelo and P.~Talavera,
  JHEP {\bf 9805} (1998) 014.


\bibitem{nosaltres2}
  I.~Rosell, J.~J.~Sanz-Cillero and A.~Pich, work in progress.

\bibitem{oneloop}
  I.~Rosell, P.~Ruiz-Femenia and J.~Portoles,
  JHEP {\bf 0512} (2005) 020.



\end{thebibliography}
\end{document}